\documentclass[aps,prb,twocolumn,showpacs,superscriptaddress]{revtex4}
\usepackage{graphicx}

\begin{document}

\title {Magnetic switching and phase competition in the multiferroic
antiferromagnet $\rm Mn_{1-x}Fe_xWO_4$ }

\author{F.~Ye}
\email{yef1@ornl.gov}
\affiliation{Neutron Scattering Science Division,
Oak Ridge National Laboratory, Oak Ridge, Tennessee 37831-6393, USA }

\author{Y.~Ren}
\affiliation{X-ray Science Division, Argonne National Laboratory,
Argonne, Illinois 60439, USA}

\author{J.~A.~Fernandez-Baca}
\affiliation{Neutron Scattering Science Division,
Oak Ridge National Laboratory, Oak Ridge, Tennessee 37831-6393 }
\affiliation{Department of Physics and Astronomy,
The University of Tennessee, Knoxville, Tennessee 37996-1200, USA}

\author{H.~A.~Mook}
\affiliation{Neutron Scattering Science Division,
Oak Ridge National Laboratory, Oak Ridge, Tennessee 37831-6393 }

\author{J. W.~Lynn}
\affiliation{NIST Center for Neutron Research,
Gaithersburg, Maryland, 20899, USA}

\author{R.~P.~Chaudhury}
\affiliation{Department of Physics and TCSUH, University of
Houston, Houston, Texas 77204-5002, USA}

\author{Y.-Q. Wang}
\affiliation{Department of Physics and TCSUH, University of
Houston, Houston, Texas 77204-5002, USA}

\author{B.~Lorenz}
\affiliation{Department of Physics and TCSUH, University of
Houston, Houston, Texas 77204-5002, USA}

\author{C.~W.~Chu}
\affiliation{Department of Physics and TCSUH, University of
Houston, Houston, Texas 77204-5002, USA}

\date{\today}

\begin{abstract}
Elastic neutron scattering is used to study the spin correlations in
the multiferroic $\rm Mn_{1-x}Fe_{x}WO_4$ with $x=0.035, 0.05$ and
$0.10$. The noncollinear, incommensurate (ICM) magnetic structure
associated with the ferroelectric (FE) phase in pure $\rm MnWO_4$ is
suppressed at $x=0.035$ and completely absent at $x=0.10$.  The ICM
spin order and FE phase can be restored by applying a magnetic field
along the spin easy-axis. The low-$T$ commensurate magnetic
structure extends in both H/T with increasing Fe concentration. The
systematic evolution of the magnetic and electric properties
indicates that the noncollinear ICM spin order results from
competing magnetic interactions and its stabilization can be tuned
by the internal ($x$) or external (magnetic field) perturbations.
\end{abstract}

\maketitle

Magnetoelectric multiferroic materials, which exhibit ferroelectric
(FE) and magnetic orders simultaneously, have attracted great
attention in recent years.
\cite{cheong07,tokura06,eerenstein06,khomskii06} A number of
multiferroic materials among transition metal oxides have been
discovered including $\rm RMnO_3$,\cite{kimura03,goto04} $\rm
TbMn_2O_5$,\cite{hur04} $\rm Ni_3V_2O_8$,\cite{lawes05} $\rm
LuFeO_4$,\cite{ikeda05} and $\rm CuFeO_2$.\cite{kimura06} The
ability to mutually control both the electric and magnetic
properties, for example, to flop the electric polarization ($P$) via
external magnetic field ($H$) (Ref.~\onlinecite{kimura03}) or to
reverse the magnetic helicity by electric field
($E$),\cite{yamasaki07} makes this family of functional materials
promising candidates for future technological applications. One
common feature of the multiferroic materials is the presence of long
wavelength magnetic structures with noncollinear spin
configurations.  Several microscopic or phenomenological mechanisms
\cite{katsura05,sergienko06prb,mostovoy06} have been proposed to
explain the spontaneous polarization induced by inhomogeneous
magnetic order $M$. Essentially, the nonlinear coupling between $P$
and $M$ becomes possible in frustrated systems with spiral or helical
magnetic structure. The symmetry allowed term $PM\partial M$ gives
rise to the electric polarization as soon as an appropriate magnetic
order sets in.

In the long-sought control of electric properties by magnetic field,
the wolframite structure $\rm MnWO_4$ appears to be a unique
magnetic material exhibiting the multiferroic behavior.
\cite{taniguchi06,heyer06,arkenbout06} Unlike the rare-earth
manganites where the stabilization of a spiral magnetic structure at
the transition metal ion site often involves the ordering of the
rare-earth spins,\cite{prokhnenko07a} $\rm MnWO_4$ is known to be a
frustrated antiferromagnet (AF) with only one kind of magnetic ion. The
Mn$^{2+}$ spins (S=5/2) in the system undergo several magnetic
transitions in zero field (H=0).\cite{lautenschlager93} The
low-temperature ($T$) magnetic structure (AF1, E-type) is collinear
and commensurate (CM) [Fig.~1(a)], with wave vector $\textbf{k} =$
($\pm$1/4,1/2,1/2). At temperatures between 7~K ($T_{N1}$) and 12~K
($T_{N2}$), the magnetic structure evolves into an incommensurate
(ICM) elliptical spiral configuration (AF2) accompanied by the
spontaneous electric polarization along the crystalline $b$-axis
[Fig.~1(b)]. Between $T_{N2}$ and $T_{N3} (\approx 13.5$~K),
MnWO$_4$ becomes collinear ICM and paraelectric (AF3). Recently, it
was found that by doping Fe at the Mn site, the electric properties
of $\rm Mn_{1-x}Fe_{x}WO_4$ ($x\approx0.10$) are considerably
modified.\cite{chaudhury08} The FE phase observed in MnWO$_4$ at H=0
is completely suppressed but can be restored with moderate magnetic
field. In addition, there appears to be transitions between low and
high electric polarized states in field.  Magnetic phase diagrams
were proposed based on a mean-field approach,\cite{chaudhury08} to
account for the observed interplay between the magnetic and electric
properties.  However, an experimental investigation of the magnetic
structure evolution and its intimate relation with the multiferroic
property induced by both doping and magnetic field is still lacking.
In this Brief Report, we report high resolution elastic neutron scattering
measurements of the magnetic correlations in multiferroic $\rm
Mn_{1-x}Fe_xWO_4$ single crystals. We observed a systematic
evolution of the phase diagram with increasing Fe doping. Under zero
magnetic field, the noncollinear, incommensurate (AF2) phase shrinks
with increasing Fe concentration and disappears completely before
$x=0.10$.  This phase is recovered with magnetic field applied along
the spin easy-axis.  Our results demonstrate that the exotic
magnetic structure results from the delicate balance between
competing magnetic interactions and are highly sensitive to small
internal and external perturbations.

\begin{figure}[ht!]
\includegraphics[width=3.2in]{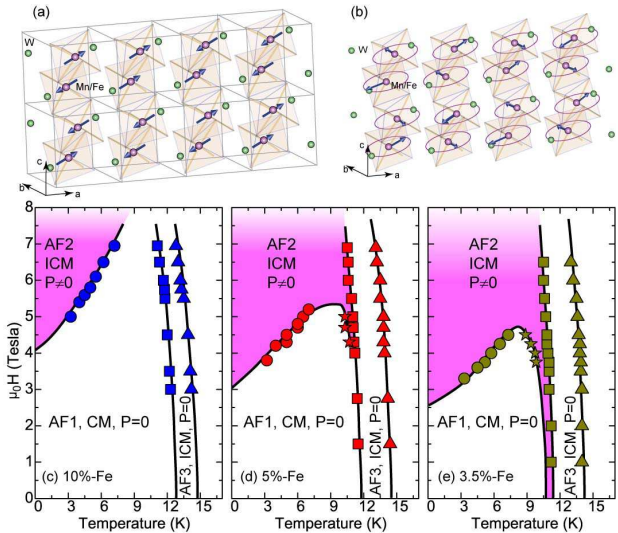}
\caption{\label{fig:structure} (Color online) (a) Crystal and
magnetic structures of $\rm Mn_{1-x}Fe_xWO_4$ in the collinear,
commensurate phase at low $T/H$ region. The magnetic spins lie in
the $ac$ plane with the moment canted to the $a$-axis about
40$^o$.\cite{powder}  (b) The magnetic structure in the
noncollinear, incommensurate phase at high magnetic field. (c)-(e)
The $H/T$ phase diagram at $\rm x=0.10$, 0.05, and 0.035.
}
\end{figure}

Single crystals of $\rm Mn_{1-x}Fe_{x}WO_4$ ($x=0.035,0.05,0.10$,
mass $=3-5$ g) were grown in an image furnace. The neutron
scattering measurements were performed at the HB1A three axis
spectrometer at the Oak Ridge National Laboratory and BT9
spectrometer at the NIST Center for Neutron Research. A closed-cycle
refrigerator in combination with a 7-T vertical field cryomagnet
was used to achieve the desired temperature and field.  The crystals
were aligned in the scattering plane defined by the two orthogonal
wave vectors (1,0,-2) and (0,1,0), in which the magnetic Bragg peaks
and other structural peaks can be surveyed. The incident neutron
energy was fixed at $14.7$~meV using pyrolytic graphite crystals as
monochromator, analyzer, and filter. 

\begin{figure}[ht!]
\includegraphics[width=3.2in]{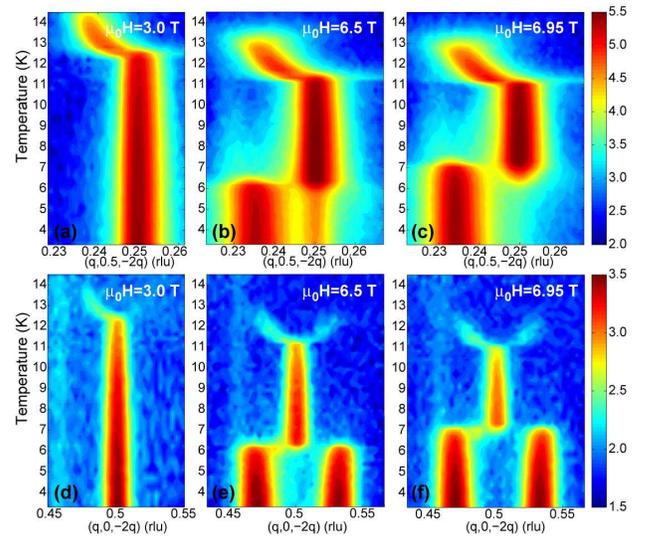}
\caption{\label{fig:qscans} (Color online) The temperature dependence of
the magnetic [panels (a)-(c)] and structural [panels (d)-(f)] orders
in $\rm Mn_{0.90}Fe_{0.10}WO_4$ at fields of $\mu_0H=3.0$, 6.5,
and 6.95~T. All measurements are done upon warming after the sample
is cooled to base temperature in the desired field.
}
\end{figure}

Figures 2(a)-2(c) show the temperature dependence of the magnetic
scattering peaks in $\rm Mn_{0.90}Fe_{0.10}WO_4$ at several magnetic
fields. At $\mu_0H=3$~T, only the CM magnetic peak
[$q_{CM}=(0.25,0.5,-0.5)$] exists at low temperatures.  The peak
intensities gradually decrease upon warming toward $\rm
T_{N1}\approx12.4$~K. Above $\rm T_{N1}$, the magnetic peak moves
abruptly to the ICM position $q_{ICM}=(\xi,0.5,-2\xi)$, with $\xi$
ranging from 0.235 to 0.25.\cite{icmnote} This is in sharp
contrast with pure MnWO$_4$, where the noncollinear, ICM magnetic
phase is present for $7<T<12$~K. The ICM phase in the doped system
is observed when the magnetic field increases.  For $\mu_0H=6.5$~T, the
ICM magnetic peaks dominate at low-$T$ with remnant CM scattering
indicative of phase competition. For a higher field
of 6.95~T, the system exhibits only the long range ICM magnetic
order at low-$T$. The $H/T$-phase diagram is constructed by
identifying the corresponding transitions, as summarized in
Fig.~1(c). The magnetic field induced structural transitions are
displayed in Figs.~2(d)-2(f). The superlattice peaks caused by the
structural distortion closely track the magnetic order. Our
observations are similar to the strong spin-lattice coupling
observed in other multiferroic materials \cite{chapon04,ye06} and
consistent with the result reported in pure MnWO$_4$.\cite{taniguchi08} 

The modification of the phase diagram at lower Fe-doping is
exemplified by studying the $T$- and $H$-dependences of the magnetic
orders in $\rm Mn_{0.95}Fe_{0.05}WO_4$. Figures 3(a) and 3(b)
display the thermal evolution of the integrated intensities of the
CM and ICM phases at several magnetic fields, respectively. At low
magnetic fields ($\mu_0H<3$~T), the ICM phase is restricted to the
high-$T$ collinear order between 11~K and 14~K. At low-$T$, only the
CM phase exists and there is no trace of the ICM magnetic state. For
3~T$<\mu_0H<5$~T, the system displays a coexistence of both phases.  The
spectral weight of the CM magnetic scattering shifts gradually to
the ICM one at higher fields. For $\mu_0H>5$~T, the low-$T$ phase is
entirely ICM and ferroelectric. We have also studied the
$H$-dependence of the magnetic scattering, as shown in
Figs.~3(c) and 3(d). The low-$T$ ICM phase stabilized under field
cooling (FC) changes abruptly into the CM phase when the magnetic
field is lowered. Consistent with the $T$-dependent measurements,
the critical magnetic field required for the ICM-CM transition
decreases as the measured temperature lowers.  For comparison, the
electric polarization measurement under field lowering at similar
temperatures is plotted in Fig.~3(f). The observed correlation
between the ICM magnetic scattering and $P$ highlights the intimate
interplay between the magnetic and the electric properties.

\begin{figure}[th!]
\includegraphics[width=3.2in]{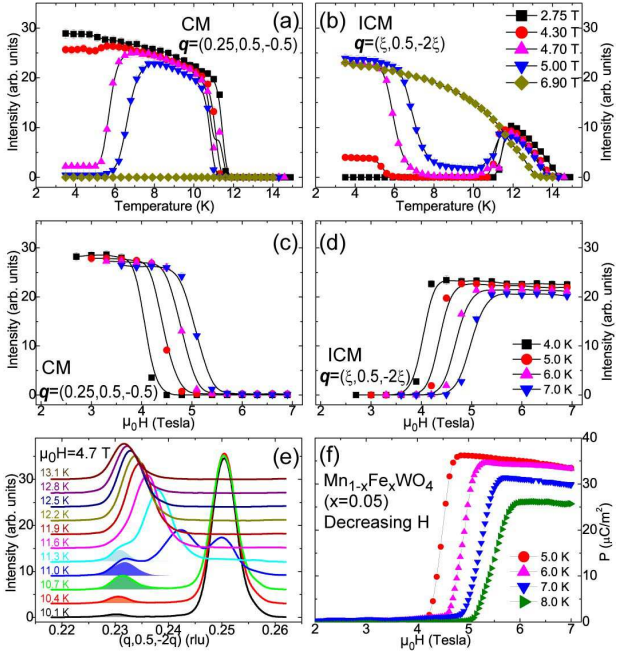}
\caption{\label{fig:5per} (Color online)
Magnetic and electric properties of $\rm Mn_{0.95}Fe_{0.05}WO_4$
(x=0.05). The $T$-dependence of the integrated intensities
of (a) the CM and (b) the ICM magnetic peaks at selective fields.
(c) and (d) shows the field dependence of integrated intensities of
the CM/ICM peaks at different temperatures. The sample is field
cooled to the desired temperature and the measurements are carried
upon lowering the field. (e) The wave vector scans near the
transition at $\mu_0H=4.7$~T. Shadow areas are the Gaussian fits to
the ICM peak. (f) The field dependence of polarization ($P$) at
different temperatures.
}
\end{figure}

Most interestingly, the $x=0.05$ sample exhibits a reentrant
behavior (the disappearance and reappearance of the low-$T$ ICM phase) in a
narrow range of magnetic fields (4~T$<\mu_0H<5$~T) just below the CM/ICM
transition. In Fig.~3(e), we show a series of wave vector scans
across both the ICM and CM magnetic Bragg peaks as the sample is
warmed at $\mu_0H=4.7$~T. Although the low-$T$ scattering profile is
dominated by the CM scattering peaks at (0.25,0.5,-0.5), the
emergence of an ICM magnetic peak is clearly seen at 10.1~K at the
wave vector $q_{ICM}=$(0.23,0.5,-0.46). The intensity initially grows
with $T$, reaches a maximum at 10.7~K, and is suppressed upon further
warming. The wave vector shows little temperature variation. At
higher $T$'s, the magnetic scattering evolves into the collinear,
ICM phase, where the peak position shows a strong temperature
dependence.

\begin{figure}[ht!]
\includegraphics[width=3.2in]{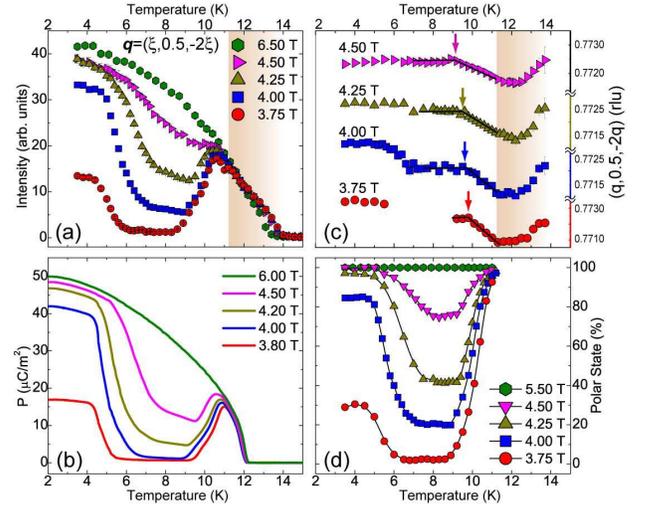}
\caption{\label{fig:3per}
(Color online)
Magnetic and electric properties of $\rm Mn_{0.965}Fe_{0.035}WO_4$
(x=0.035). (a) $T$-dependence of the ICM magnetic peaks at
different magnetic fields. (b) The polarization measurements under
the same thermal protocol. (c) The evolution of the ICM wave vectors
in selective magnetic fields. The arrows mark the transition where
the system reenters the ferroelectric phase. (d) The $T$ dependence
of the polar state volume fraction at different fields.
Uncertainties in the plots are statistical and represent one
standard deviation.
}
\end{figure}

The reentrant ferroelectric behavior is more evident at $x=0.035$.
Figure 4(a) displays the $T$ dependence of the integrated intensities
for this sample extracted from the ICM magnetic peak at several
$H$'s.  At $\mu_0H=3.75$~T, the low-$T$ ICM intensity collapses
suddenly near 5~K, accompanying the enhancement of the CM peak (data
not shown). Such a transition between the competing magnetic phases
is similar to the low field data at $x=0.05$. However, the ICM
intensity starts to increase again as $T$ just crosses 9~K, which is
about 2~K lower than the expected AF2/AF3 phase boundary. The latter
increase in the ICM intensity is accompanied by the reappearance of
the FE polarization [Fig.~4(b)] suggesting that the ICM magnetic
state between 9 and 12~K is the same as the low-$T$ noncollinear,
electrical polar phase.

Unlike the case of higher Fe doping where the boundary of the
paraelectric phase is well defined, the transition from low-$T$ ICM
phase to the high-$T$ ICM phase becomes less discernible at
$x=0.035$.  To better characterize the phase boundary, we tracked the
thermal evolution of the ICM peak position at different fields with
higher resolution [Fig.~4(c)].  Although the temperature shift is
much smaller compared to those presented in Figs.~2(a)-2(f) and
3(e), the boundaries at individual AF1/AF3 transitions are
visible despite the lack of an anomaly in intensity [Fig.~4(a)]. In
addition, the wave vector shows a distinct variation when the system
re-enters the FE/ICM phase. The boundaries obtained from neutron
scattering are in agreement with the polarization data in Fig.~4(b).
Finally, we estimate the volume percentage $V_P(\%)$ of the polar
state [Fig.~4(d)]. This is done by calculating
$V_P(\%)=I_{ICM}/(I_{ICM}+I_{CM})$, where $I_{ICM}$ and $I_{CM}$ are
the integrated ICM/CM intensities. Although there is considerable
non-polarized volume fraction at lower $H/T$ regime, the fraction of
the FE phase increases quickly with field and becomes saturated at
$\mu_0H=4$~T. We further notice that the magnetic fluctuations below 12~K
are mainly incommensurate, but the CM spin order gradually builds up
at lower temperatures.

It is known that the magnetic properties of frustrated spin systems
are sensitive to impurities and external stimuli. In contrast to the
geometrically frustrated antiferromagnet CuFeO$_2$ where a 1.2\% Al
doping at the Fe site stabilizes the multiferroic phase,\cite{kanetsuki07} 
the substitution of Mn by Fe in MnWO$_4$
suppresses such phase. The electric polar state observed in a
moderate temperature range at $x=0$ is completely absent at
$x=0.10$. Instead of forming an $E$-type spin order with zigzag AF
chains along the $c$-axis in MnWO$_4$, pure FeWO$_4$ has a
relatively simple magnetic structure with magnetic wave vector
$q_M=(0.5,0,0)$.\cite{stuber01} Previous inelastic neutron
scattering (INS) studies on MnWO$_4$ reported that the spin wave
dispersion relations can only be described by including magnetic
coupling up to ninth nearest neighbors.\cite{ehrenberg99} The INS
data demonstrate that the stabilization of the CM spin order
requires long range magnetic interactions. It is expected that the
substitution leads to modifications of the local environment around
the magnetic ions, which ultimately causes a redistribution of
exchange coupling strengths. On the other hand, despite the complex
interactions associated with the crystal structure, Chaudhury {\it
et al.}\cite{chaudhury08} proposed a simple mean-field model,
including nearest and next-nearest-neighbor interactions, spin
anisotropy ($K$), and external magnetic field, to explain the
observed phase diagram in Fe-doped MnWO$_4$. It was speculated that
the lack of ferroelectricity (at H=0) in Mn$_{0.9}$W$_{0.1}$WO$_4$ is
due to the increase in the uniaxial anisotropy $K$. Based on a
pressure-dependent study on MnWO$_4$,\cite{chaudhury07} which
showed a pressure-induced suppression of the FE phase, the Fe-doping
can be viewed as an internal (chemical) pressure effect on the
structure. Indeed, a noticeable decrease in Mn-O bonding distance
within the edge-sharing octahedron is observed in the neutron powder
refinement for the $x=0.10$ sample.

In summary, our neutron scattering studies reveal a systematic
evolution of magnetic properties in the multiferroic $\rm
Mn_{1-x}Fe_xWO_4$. The ICM spin structure associated with the FE
phase at H=0 shrinks with increasing Fe concentration. The
noncollinear magnetic configuration in the doped system is restored
with the application of external magnetic field. Those results
indicate that the complex spin structure accompanying spontaneous
electric polarization results from competing magnetic interactions
and is subject to both internal (chemical doping) and external
(magnetic field) perturbations.

We are grateful to R. S Fishman and T. Kimura for helpful
discussions. This work was partially supported by Division of
Scientific User Facilities of the Office of Basic Energy Sciences,
US Department of Energy. This work utilized facilities supported in
part by the National Science Foundation under Agreement No.
DMR-0454672.  Work at Houston was supported by the T.L.L. Temple
Foundation, the J.J. and R. Moores Endowment, the U.S.  Air Force
Office of Scientific Research, and the State of Texas through TCSUH.


\begin{references}
\bibitem{cheong07} S.-W. Cheong and M. Mostovoy, Nature Materials
{\bf 6}, 13 (2007).
\bibitem{tokura06} Y. Tokura, Science {\bf 312} 1481 (2006).
\bibitem{eerenstein06} W. Eerenstein, N. D. Mathur, J. F. Scott,
Nature {\bf 442}, 759 (2006).
\bibitem{khomskii06} D. I. Khomskii, J. Magn. Magn. Mater. {\bf
306}, 1 (2006).
\bibitem{kimura03} T. Kimura, T. Goto, H. Shintani, T. Arima, and
Y.~Tokura, Nature (London) {\bf 426}, 55 (2003).
\bibitem{goto04} T. Goto, T. Kimura, G. Lawes, A. P. Ramirez, and Y.
Tokura, Phys. Rev. Lett. {\bf 92}, 257201 (2004).
\bibitem{hur04} N. Hur, S. Park, P.~A.~Shama, J.~S.~Ahn, S.~Guha,
and S.-W.~Cheong, Nature (London) {\bf 429}, 392 (2004).
\bibitem{lawes05} G. Lawes, A.~B.~Harris, T.~Kimura, N.~Rogado,
R.~J. Cava, A. Aharony, O. Entin-Wohlman, T. Yildirim, M.
Kenzelmann, C. Broholm, and A.~P. Ramirez, Phys. Rev. Lett. {\bf
95}, 087205 (2005).
\bibitem{ikeda05} N. Ikeda, H. Ohsumi, K. Ohwada, K. Ishii, T.
Inami, K. Kakurai, Y. Murakami, K. Yoshii, S. Mori, Y. Horibe, and
H. Kit\^{o}, Nature (London), {\bf 436} 1136 (2005).
\bibitem{kimura06} T. Kimura, J. C. Lashley, and A. P. Ramirez,
Phys. Rev. B {\bf 73}, 220401(R) (2006).
\bibitem{yamasaki07} Y. Yamasaki, H. Sagayama, T. Goto, M. Matsuura,
K. Hirota, T. Arima, and Y. Tokura, Phys. Rev. Lett. {\bf 98},
147204 (2007).
\bibitem{katsura05} H. Katsura, N. Nagaosa, and A. V. Balatsky,
Phys.  Rev. Lett. {\bf 95}, 057205 (2005).
\bibitem{sergienko06prb} I. A. Sergienko and E. Dagotto, Phys. Rev.
B {\bf 73}, 094434 (2006).
\bibitem{mostovoy06} M. Mostovoy, Phys. Rev. Lett. {\bf 96}, 067601
(2006).
\bibitem{taniguchi06} K. Taniguchi, N. Abe, T. Takenobu, Y. Iwasa,
and T. Arima, Phys. Rev. Lett. {\bf 97}, 097203 (2006).
\bibitem{heyer06} O. Heyer, N. Hollmann, I Klassen, S. Jodlauk, L.
Bohat\'{y}, P. Becher, J.~A. Mydosh, T. Lorenz, and D. Khomskii, J.
Phys.: Condens. Matter {\bf 18}, L471 (2006).
\bibitem{arkenbout06} A. H. Arkenbout, T. T. M. Palstra, T.
Siegrist, and T. Kimura, Phys. Rev. B {\bf 74}, 184431 (2006).
\bibitem{prokhnenko07a} O. Prokhnenko, R. Feyerherm, E. Dudzik, S.
Landsgesell, N. Aliouane, L.~C.~Chapon, and D.~N.~Argyriou, Phys.
Rev.  Lett. {\bf 98}, 057206 (2007), O. Prokhnenko, R.~Feyerherm, M.
Mostovoy, N. Aliouane, E. Dudzik, A.~U.~B. Wolter, A. Maljuk, and D.
N. Argyriou, {\it ibid} {\bf 99}, 177206 (2007).
\bibitem{lautenschlager93} G. Lautenschl\"{a}ger, H. Weitzel, T.
Vogt, R. Hock, A. B\"{o}hm, M. Bonnet, and H. Fuess, Phys.  Rev. B
{\bf 48}, 6087 (1993). H. Ehrenberg, H. Weitzel, C. Heid, H. Fuess,
G. Wltschek, T. Kroener, J. van Tol, and M. Bonnet, J.
Phys.: Condens. Matter {\bf 9}, 3189 (1997).
\bibitem{chaudhury08} R. P. Chaudhury, B. Lorenz, Y. Q. Wang, Y. Y.
Sun, and C. W. Chu, Phys.  Rev. B {\bf 77}, 104406 (2008).
\bibitem{powder} High resolution neutron powder diffraction
measurements confirm the low-$T$ magnetic structure of $x=0.10$
sample at AF1 phase is the same as the pure MnWO$_4$. 
\bibitem{icmnote} Although the actual wave vector associated with
the ICM magnetic peak is $(h,k,l)=(0.234,0.5,-0.46)$, the current
scattering configuration still allows us to probe the ICM peak due
to the coarse vertical resolution of the spectrometer. 
\bibitem{chapon04} L. C. Chapon, G. R. Blake, M. J. Gutmann, S.
Park, N. Hur, P. G. Radaelli, and S.-W. Cheong, Phys. Rev. Lett.
{\bf 93}, 177402 (2004). 
\bibitem{ye06} F. Ye, Y. Ren, Q.~Huang, J.~A.~Fernandez-Baca,
Pengcheng Dai, J.~W.~Lynn, and T.~Kimura, Phys. Rev. B {\bf 73},
220404(R) (2006).
\bibitem{taniguchi08} K. Taniguchi, N. Abe, H. Sagayama, S. Otani,
T. Takenobu, Y. Iwasa, and T. Arima, Phys. Rev. B {\bf 77}, 064408 (2008).
\bibitem{kanetsuki07} S. Kanetsuki, S. Mitsuda, T. Nakajima, D.
Anazawa, H. A. Katori, and K. Prokes, J. Phys.: Condens. Matter
{\bf 19}, 145244 (2007).
\bibitem{stuber01} N. St\"{u}\ss er, Y. Ding, M. Hofmann, M.
Reehuis, B. Ouladdiaf, G. Ehlers, D. G\"{u}nther, M. Mei\ss ner, and M.
Steiner, J. Phys.: Condens.  Matter{\bf 13}, 2753 (2001).
\bibitem{ehrenberg99} H. Ehrenberg, H. Weitzel, H. Fuess, and B.
Hennion, J. Phys.: Condens.  Matter {\bf 11}, 2649 (1999).
\bibitem{chaudhury07} R.~P.~Chaudhury, F.~Yen, C.~R.~dela Cruz, B.
Lorenz, Y.~Q.~Wang, Y.~Y.~Sun, and C.~W.~Chu, Physica B {\bf 403},
1428 (2007).
\end{references}
\end{document}